# Damage accumulation induced metal-insulator transition through ion implantation of ScN thin films

Charlotte Poterie,[1*] Marc Marteau,[1] Per Eklund,[2,3] Thierry Cabioc'h[1], Jean-François Barbot[1] Arnaud le Febvrier[2**]

[1] PPRIME Institute, CNRS, Université de Poitiers–ENSMA, UPR 3346, SP2MI, TSA 41123, 86073 Poitiers cedex 9, France
[2] Inorganic Chemistry, Department of Chemistry - Ångström, Uppsala University, Box 538, SE-751 21 Uppsala, Sweden
[3] Thin Film Physics Division, Department of Physics, Chemistry and Biology (IFM), Linköping University, SE-581 83 Linköping, Sweden

Corresponding authors:
*charlotte.poterie@univ-poitiers.fr
**arnaud.lefebvrier@kemi.uu.se

**ABSTRACT**. Ion implantation is a powerful approach for tuning the electrical properties of materials through controlled doping and defect engineering, with applications in thermoelectrics and microelectronics. Scandium nitride (ScN) is particularly sensitive to irradiation-induced disorder, with transport properties spanning several orders of magnitude and multiple conduction mechanisms involved. In this study, we investigate the evolution of electrical transport in epitaxial ScN thin films undergoing accumulated irradiation damage at an initial defect state. A phenomenological damage-accumulation model was successfully combined with temperature dependent resistivity and Hall effect measurements to elucidate the impact of defect buildup on electrical transport and to provide physically grounded, quantitative insight into the nature and accumulation of irradiation-induced defects. It reveals two distinct defects-generation regimes of electrically active defects. At low doses, direct-impact damage produces stable and isolated acceptor-type complex defects, ($V_{Sc}$–X) with $V_{Sc}$ a scandium vacancy and X denoting residual impurities, leading to a gradual increase in resistivity. At higher doses, defect accumulation dominates through a multi-hit process, giving rise to point-defect buildup and carrier localization, resulting in hopping-dominated transport. This localized regime is thermally unstable and recovers upon low-temperature annealing. We further demonstrate that the residual defect landscape strongly influences both the critical dose for the metal–insulator transition and the localization strength: films grown on $Al_2O_3$ exhibit an earlier transition and weaker localization than those grown on MgO. These results highlight ion implantation as an effective route for engineering disorder-induced localization in ScN, with the initial film quality playing a decisive role.

## I. INTRODUCTION

Ion implantation is a key process for the introduction of dopants in semiconductor materials. One of the critical secondary effects of this technique is the damage induced in the crystalline structure by collisions between the incoming ions and the target lattice. These collisions lead to atomic displacements and the formation of defects, which can significantly modify the electrical properties of the material by introducing energy levels, altering carrier concentration, mobility, and conduction mechanisms. Beyond its role in doping, the impact of irradiation-induced defects on the performance and reliability of modern devices has motivated the development of so-called *defect engineering* strategies, in which controlled disorder is intentionally introduced to tailor material properties. Ion irradiation provides a controlled and versatile approach to introduce defects, such as point defects (vacancies, interstitials, substitutions, antisites) and more extended defects (cluster of point defects, dislocations…) and to investigate their impact on charge transport. Thereby, it has received significant attention for thermoelectric applications due to its possibilities in decreasing the thermal conductivity while using the doping effects to control the electrical properties. [1–5] Moreover, recent studies have demonstrated that ion implantation enables the decoupling of the Seebeck coefficient from electrical conductivity, a key requirement for enhancing thermoelectric performance. [6,7]

Scandium nitride (ScN) is a transition-metal nitride that has attracted increasing interest due to its combination of high electrical conductivity, thermal stability, and compatibility with epitaxial growth on common oxide substrates. In its as-deposited state, ScN typically exhibits degenerate *n*-type metallic behavior, originating from native point defects and unintentional impurities such as oxygen. [8–17] These characteristics make ScN a promising material for applications in thermoelectrics, electronics, and radiation-tolerant devices. [13]

Ion implantations of noble gases (Ar and He), Li, and Mg in ScN thin films highlight that the thermal conductivity can be reduced due to increased defects scattering. [1–3,18] These defects also act as traps for the carriers, inducing an increase in the Seebeck coefficient. [1–3,18–20] In turn, the electrical resistivity is increased and a change in the conduction mode is usually observed from metallic to semiconducting conduction. [1,3,19–21] This semiconducting mode has been ascribed to a hopping conduction due to the defects inducing localized states. Interestingly, this localization occurs both for noble gases [3,21] which are inert and for chemically active elements such as Mg, O and C. [1,19,20] This change in the conduction mode has also been observed in ScN thin films while doping with Mg during deposition and was ascribed to a semiclassical Anderson localization. [22] In this case, the films exhibit no structural disorder due to the use of high deposition temperature but the localization of the carriers occurred due to the inhomogeneous distribution of charged dopants inducing potential fluctuations. This metal to insulator transition (MIT) was also observed as a function of deposition temperature using high power impulse magnetron sputtering (HiPIMS). [23] In this work, the MIT was attributed to the dislocation's density increasing for lower deposition temperatures along with a rhombohedral distortion. These works highlight the effect of defects in the control of the electrical properties.

*charlotte.poterie@univ-poitiers.fr
**arnaud.lefebvrier@kemi.uu.se

Several explanations for the MIT have been proposed. In general, it occurs according to Mott principle through increased Coulomb interactions between carriers [24] or according to Anderson principle through disorder which induce incoherent potential fluctuation. [24,25] Mott considered that there is a critical carrier concentration ($n_c = (\frac{0.25}{a_B^*})^3$, with $a_B^*$ the effective Bohr radius) below which the Coulomb interactions become large enough to induce localization of the carriers. In the case of ScN, assuming an effective mass of $0.6m_0$ [26], we estimate the critical carrier concentration to be $n_c=2\times10^{19}$ $cm^{-3}$. Transitions occurring from a combine effect of these two principles is possible as defects have both an effect on carrier concentration and on structural quality. Studies have shown that tuned localization could be used for improving TE properties. [27–29] Selective Anderson localization for minority charge carriers was reported as an enhanced strategy to improve the Seebeck coefficient and decrease the lattice thermal conductivity while maintaining the electrical conductivity. [27,28] A transport model, verified on GeSbTe, has also been developed and has predicted an increase of 20% in the maximum zT by tuning both the doping and the structural disorder. [29]

Understanding carrier localization and the metal – insulator transition (MIT) is crucial for controlling electrical transport. Localization is commonly quantified by the Mott temperature $T_M$, which reflects the energy scale for hopping between localized states. Reported $T_M$ values in ScN thin films strongly depend on the implanted ion species and damage level, indicating that irradiation conditions critically influence localization strength. [3,19,21,30] However, direct comparison between studies remains challenging due to differences in deposition conditions (substrates, temperature, Ar/N2 ratio, …) resulting in different microstructure and residual defect states. Recent irradiation studies on related nitrides have further shown that the MIT occurs at a critical damage level and that both the transition threshold and localization strength depend on the initial crystalline state, underscoring the key role of the pristine film quality. [6]

In this work, we investigate the evolution of electrical transport in epitaxial ScN films grown on MgO (001) and Al$_2$O$_3$ (0001) substrates under controlled ion irradiation, spanning a wide range of accumulated damage. By combining temperature-dependent resistivity and Hall-effect measurements with phenomenological damage-accumulation models, we identify distinct regimes of defect formation and conduction behavior. We show that irradiation drives a progressive transition from metallic transport to defect-induced localization via hopping conduction, with the transition dose, localization strength, and saturation behavior strongly dependent on the substrate. Our results using the damage-accumulation model that we adapted for quantifying electrically active defects provide new insight into the interplay between film quality, defect accumulation and carrier localization in ScN. Therefore, it establishes irradiation as an effective tool for tuning transport properties in nitride semiconductors.

## II. METHODS

DC magnetron sputtering in an ultra-high vacuum chamber was used for depositing epitaxial-like ScN thin films on different substrates (MgO (001) or Al$_2$O$_3$ c-cut) achieving the following orientation relationship: $[001](001)ScN \parallel [001](001)$ MgO and $[1\bar{1}0](111)ScN \parallel [11\bar{2}0](0001)$ Al$_2$O$_3$. [11,13]
The double side polished MgO (001) substrates were cleaned beforehand with detergent steps (described elsewhere [31]) followed by a 10 min ultrasonic bath of acetone, then of ethanol, and finally blown dry with a N$_2$-gun. The one-side polished Al$_2$O$_3$ c-cut substrates were simply cleaned using acetone, ethanol and then dried with a N$_2$-gun. The films were deposited in a ultra-high vacuum chamber with a $10^{-7}$ Pa base pressure (more details on the deposition chamber can be found elsewhere [32]). In both cases, a 50 mm (2-inch) Sc target (Mateck, 99.95%) was operated with a power of 110 W. The working pressure was kept at 0.27 Pa (2 mTorr) during deposition with a sputtering gas mixture of 70% Ar/30% N$_2$. The films were deposited under constant rotation at a temperature of 850°C ensuring a high crystalline quality [11], thus limiting oxidation of the films. [33]

The samples (consisting of film/substrate (MgO (001) or Al$_2$O$_3$ c-cut)) were implanted cumulatively with oxygen ions (O$^{2+}$) at room temperature (RT) and accelerated at 180 keV in the EATON VN3206 implanter, under an approximate vacuum of $1\times10^{-6}$ mbar. The SRIM-2013 software was used in full-damage-cascade mode [34] to theoretically predict the depth profile of oxygen ions implanted into the targets and determine the fluences in order to obtained the desired damage levels (called displacements per atoms (dpa)). A theoretical density of 4.27 g cm$^{-3}$ for the ScN films was used. FIG 1. illustrates the case of an estimated damage level of 0.1 dpa where the implanted oxygen concentration in the film remains low, below $1 \times 10^{-3}$ at.%. In the present study, we consider only the damage induced in the films and not in the substrates as MgO and Al$_2$O$_3$ are insulator and do not have contribution in the electrical properties measurement. Using O$^{2+}$ accelerated at 180 keV, the ions pass through the film, inducing damage in their vicinity, with the maximum damage occurring in the substrate. While the film experiences non-negligible irradiation-induced damage, almost no implanted oxygen impurities remain within its thickness. This experimental approach allows us to investigate the effect of irradiation-induced defects with a limited contribution from implanted oxygen ions. The fluences were then increased cumulatively, step by step, up to damage levels reaching 2.25 dpa for MgO and 2.15 dpa for Al$_2$O$_3$ substrates and a cumulative amount of O around $2 \times 10^{-2}$ at.%.

*charlotte.poterie@univ-poitiers.fr
**arnaud.lefebvrier@kemi.uu.se

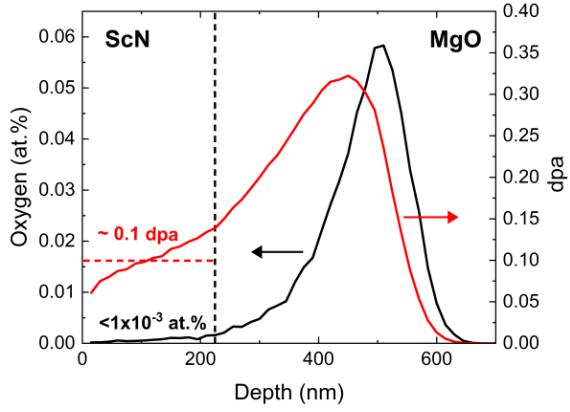

FIG 1. Oxygen concentration (black curve) and damage level (red curve) in the ScN/MgO system as a function of implantation depth for a fluence of $8\times10^{14}$ cm$^{-2}$. The dpa value in the film is considered to be roughly constant across the film thickness, with a mean value of 0.1 dpa. The concentration of implanted oxygen in the film remains below $5 \times 10^{-4}$ at.%.

After each implantation step (from 0.005 dpa to 2.25 dpa), systematic macroscopic in-plane resistivity $\rho(T)$ and Hall carrier concentration $n_H(T)$ of the films were measured between 80 and 300 K using the liquid-nitrogen cryostat of the ECOPIA HMS-5500 system, which combines the van der Pauw method with Hall-effect measurements. The mobility $\mu(T)$ was then calculated using $\mu(T) = \frac{1}{n_H(T)\rho(T)e}$, with $e$ the elementary charge. A constant magnetic field of 0.580 T was applied for all Hall measurements. For one selected implanted sample, annealing was performed at 873 K in a dedicated vacuum furnace, under a vacuum of $5\times10^{-6}$ Pa. In a non-cumulative approach and for comparison, one sample was implanted to reach directly the dose of 0.4 dpa followed by *in situ* annealing, performed up to 750 K, using the high-temperature module of the ECOPIA HMS-5500.

## III. RESULTS
### A. Substrate effects on as-grown transport

The in-plane resistivity and mobility as a function of temperature for the ScN reference samples on MgO (blue) and on Al$_2$O$_3$ (green) are shown in FIG 2. ScN thin films are usually n-type degenerate due to the incorporation of oxygen during deposition. [8–12,14] The extracted carrier concentrations, which remain constant with temperature confirm the n-type degenerate state: $9.0(5) \times 10^{20}$ cm$^{-3}$ for ScN/MgO and $8.2(5) \times 10^{20}$ cm$^{-3}$ for ScN/Al$_2$O$_3$ (in both cases, $n>n_c$). Both resistivity–temperature curves show a similar trend, exhibiting a positive slope with increasing temperature (d$\rho$/dT > 0), i.e., characteristic of metallic conduction. This metallic behavior can be described for T>T$_0$=175 K by the following equation:

$$\rho(T) = \alpha(T - T_0) + \rho_R + \rho_D(dpa) \quad (1)$$

with α describing the slope of the curve, $\rho_R$ the residual electrical resistivity and $\rho_D$ the additional electrical resistivity induced by implantation induced defects. In the reference state, $\rho_{D,0\,dpa} = 0$. The values of $\alpha$ and $\rho_R$ are reported Table 1. At low temperature, this metallic behavior can be also described using the Bloch-Grüneisen model, which considers the electron-phonon interactions:

$$\rho(T) = (\rho_R + \rho_D) + A(\frac{T}{\theta_D})^n \int_0^{\frac{\theta_D}{T}} \frac{t^n}{(e^t-1)(1-e^{-t})} dt \quad (2)$$

where A is a constant, $\theta_D$ is the Debye temperature, and $n$ a factor that depends on the interaction's nature. In this case, the curves (FIG 2(a)) were fitted using $n = 5$, which correspond to electrons scattering by acoustic phonons. [35] Generally, $\theta_D$ is also used to describe the stiffness of a material as it is related to the maximal phonon frequency $\omega_D$: $\theta_D = \frac{\hbar\omega_D}{k_B}$ (with $\hbar$ the reduced Planck constant and $k_B$ the Boltzmann constant). The extracted values of $\theta_D$ are reported Table 1. The sample grown on MgO exhibits a slightly higher $\theta_D$ than that grown on Al$_2$O$_3$, indicating that the ScN film is stiffer when deposited on MgO. This difference can be attributed to the distinct epitaxial relationships with the two substrates: the film grown on Al$_2$O$_3$ contains a higher density of residual defects, defects, which lowers the maximum phonon frequencies $\omega_D$ and therefore $\theta_D$. This difference in residual defect density is also reflected in the higher residual resistivity of the film on Al$_2$O$_3$, which in turn accounts for its lower carrier mobility, as shown in FIG 2(b).

*charlotte.poterie@univ-poitiers.fr
**arnaud.lefebvrier@kemi.uu.se

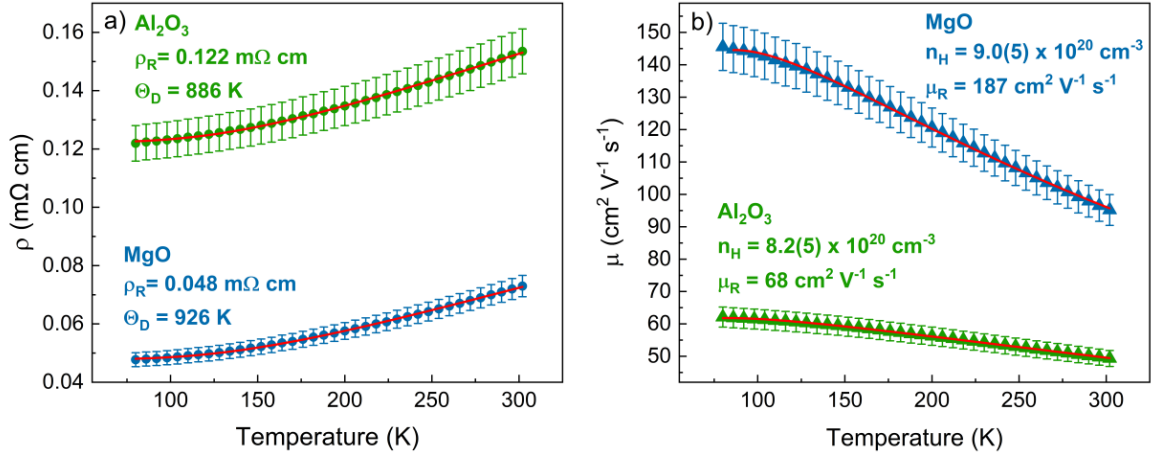

FIG 2. The temperature dependence of a) the resistivity ρ and b) the mobility μ for the reference films. In blue, the film deposited on MgO and in green the one deposited on Al$_2$O$_3$. The red dotted lines represent the Bloch-Gruneisen model (equation 2) for the resistivity (a) and, the Matthiessen law (equation 3) for the mobility (b).

Using Matthiessen's rule, the different contributions to the mobility can be separated to highlight the role of defects beyond the intrinsic lattice contribution $\mu_{lat}$. In the reference state, scattering from residual defects $\mu_R$ must be taken into account, as well as a component for scattering at grain boundaries $\mu_{GB}$, more dominant at low temperature. After implantation, an additional term $\mu_D(dpa)$, associated with irradiation-induced defects, is introduced. Accordingly, the temperature dependence of the mobility was fitted using:

$$\frac{1}{\mu(T)} = \frac{1}{\mu_{lat}(T)} + \frac{1}{\mu_R} + \frac{1}{\mu_{GB}(T)} + \frac{1}{\mu_{D(dpa)}} \quad (3)$$

With $\mu_{lat} = \mu_0 T^{-1.5}$ and $\mu_{GB} \sim \mu_1 \exp\left(\frac{-E_a}{k_B T}\right)$

where $E_a \sim 10$ meV, and $\mu_0$ and $\mu_1$ were adjusted to obtain good agreement with the experimental curves. The activation energy $E_a$ is found to be independent of the substrate, indicating that grain boundaries have a minimal impact on mobility scattering. In contrast, the extracted values of $\mu_R$ (Table 1) reveal a higher concentration of residual defects in the film deposited on Al$_2$O$_3$, as shown by its significantly lower $\mu_R$ compared with the film on MgO. This difference primarily arises from growth variants in ScN deposited on Al$_2$O$_3$, [2,3,11,30,36] which disrupt the crystal symmetry and contribute to the distinct residual defect landscape in each film. Film orientation (001 on MgO vs. 111 on Al$_2$O$_3$) is not expected to play a major role: ScN is nearly elastically isotropic (Zener anisotropy η=1.04). [37] Thus, the dominant factor governing the differences in transport properties is the residual defect landscape, not crystallographic orientation. In summary, the distinct transport behaviors of the two reference samples originate from their different epitaxial growth: the film on Al$_2$O$_3$ exhibits a higher residual defect concentration and softer bonding compared with the film grown on MgO.

TABLE I. Residual resistivity $\rho_R$, resistivity slope α, Debye temperature $\Theta_D$ and residual mobility $\mu_R$ of the reference film deposited either on MgO (001) or on c-cut Al$_2$O$_3$.

| Substrate used | $\rho_R$ (mΩ cm) | α (mΩ cm K$^{-1}$) | $\Theta_D$ (K) | $\mu_R$ (cm² V$^{-1}$ s$^{-1}$) |
|---|---|---|---|---|
| MgO | 0.048 | 1.5×10$^{-4}$ | 926 | 187 |
| Al$_2$O$_3$ | 0.122 | 1.8×10$^{-4}$ | 886 | 68 |

### B. Damage-Accumulation-Driven Evolution of Electrical Resistivity

The films deposited on MgO and Al$_2$O$_3$ were cumulatively implanted using oxygen ions. It produces electrically active defects which induce disorder in the ScN films. These defects lead to modifications of the transport properties which were determined between each implantation. The change in the electrical transport properties is assumed to be proportional to the density of electrically active defects produced and surviving *in situ* recombination during implantation. The evolution of the resistivity at RT as a function of dpa for the film deposited on MgO (001) is shown FIG 3(a) for cumulative implantation spanning from 0 (reference) to 2.25 dpa. The resistivity increases by nearly four orders of magnitude when the level of dpa reaches a maximum of 2.25 dpa, with $\rho_{2.25\,dpa,\,300K}$ = 286 mΩ cm. The variation of the electrical resistivity, is non-linear and is plotted on a logarithmic scale to highlight two distinct regimes, with a transition occurring around 0.8 dpa. In the low-damage regime, the resistivity rises sharply, whereas at higher damage levels (dpa > 0.8 dpa) the increase becomes more gradual and tends toward saturation. A pronounced increase in electrical resistivity has previously been reported in ion-implanted ScN thin films and is generally accompanied by a change in the conduction mechanism. [1,3,19–21]

*charlotte.poterie@univ-poitiers.fr
**arnaud.lefebvrier@kemi.uu.se

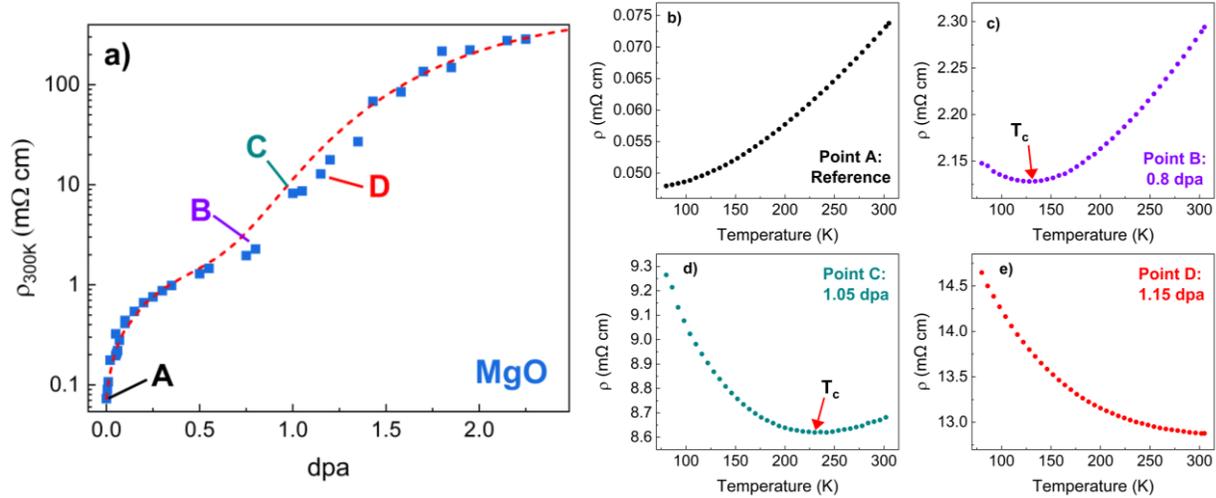

FIG 3. (a) – The evolution of the resistivity at 300K (in log-scale) of the ScN film deposited on MgO 001 substrate as a function of damage accumulation represented by the dpa. The dotted line represents the model from equation (6) discussed section 3. (b, c, d, e) – Temperature dependence of the resistivity having specific dpa: Point A) for the reference, Point B) for 0.8 dpa, Point C) for 1.05 dpa and Point D) for 1.15dpa. The uncertainty on the resistivity measurement is ± 2%.

The conduction mechanism was then investigated using temperature-dependent resistivity measurements. The reference film on MgO, labeled A in FIG 3(b)., exhibits a positive slope ($\frac{d\rho}{dT} > 0$), indicating metallic behavior (FIG 3(b), point A), with a constant carrier concentration of $9.0(5) \times 10^{20}$ cm$^{-3}$. As damage accumulates, the metallic behavior is maintained up to approximately 0.8 dpa. At this point (FIG 3(c), point B), the low-temperature slope of resistivity begins to change sign, becoming negative ($\frac{d\rho}{dT} < 0$), indicating a transition in the electrical conduction mechanism toward a semiconducting behavior below a temperature of 128 K. This temperature defined as $T_c$ is the temperature at which the slope changes from negative to positive, thus highlighting the transition from a semiconducting to a metallic behavior. $T_c$ is increasing as damage accumulates from 128 K at 0.8 dpa to 230 K at 1.05 dpa (FIG 3(d), point C). This rise of $T_c$ with increasing damage demonstrates that the semiconducting behavior becomes progressively more dominant over this temperature range as defects accumulate in the film. This accumulation gives rise to increasing disorder which induces progressive carrier localization. After reaching a damage level of 1.15 dpa (FIG 3(e), point D), the film exhibits fully semiconducting behavior across the entire measured temperature range. The analysis of the conduction modes thus reveals that the ScN film undergoes a defect-induced metal–insulator transition (MIT) over the same dpa range as the transition between the two electrical resistivity regimes observed FIG 3(a). These results indicate that the change in conduction mode is a continuous process that progressively develops once a critical damage level is reached. This is in good agreement with recent work on strain-engineered ScN films grown at different temperatures, where a MIT was observed as the growth temperature decreased, leading to increased deformation and defect density (including dislocations). [23]

*charlotte.poterie@univ-poitiers.fr
**arnaud.lefebvrier@kemi.uu.se

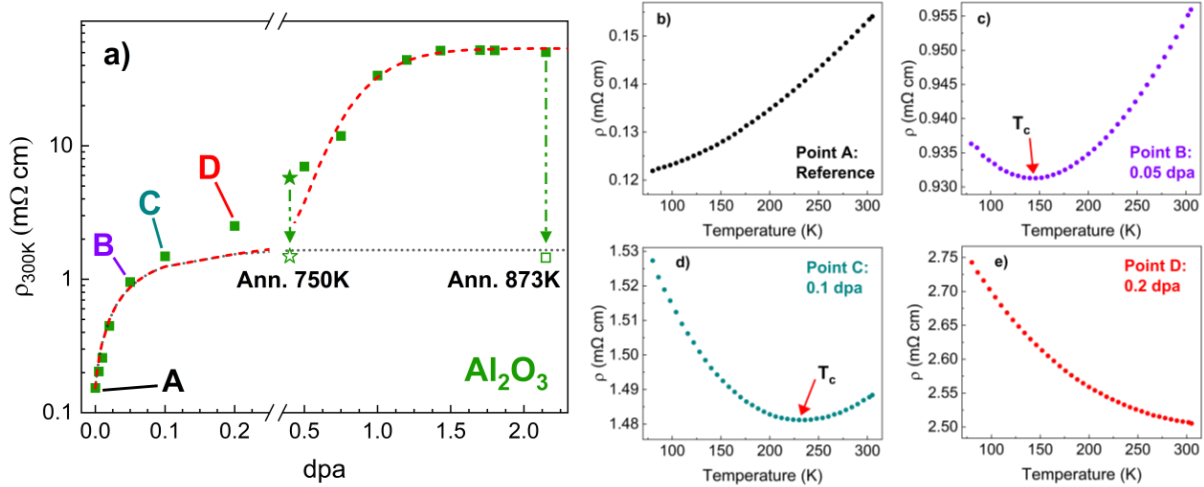

FIG 4. (a) – Evolution of the electrical resistivity at 300 K (log-scale) of the ScN film grown on c-cut $Al_2O_3$ as a function of accumulated damage (dpa). The red dotted line corresponds to the model described by Eq. (6) and discussed in Sec. 3. The filled star represents a film implanted once at 0.4 dpa, demonstrating that cumulative and single-step implantations lead to similar resistivity values. Two samples implanted in regime II were subsequently annealed, showing recovery toward regime I. (b, c, d, e) Temperature-dependent resistivity measurements highlighting the evolution of critical transport regimes: Point A) for the reference film, Point B) for 0.05 dpa, Point C) for 0.1 dpa, and Point D) for 0.2 dpa. $T_c$ represent the temperature at which the temperature at which the film undergoes a transition from semiconducting to metallic behavior. The uncertainty in resistivity measurements is ±2%.

A similar two-regime behavior of the electrical resistivity with increasing generated damage is also observed for ScN films grown on $Al_2O_3$ substrates (see FIG 4(a)). However, differences, compared to the film deposited on MgO, are observed in both the amplitude of electrical resistivity change and level of damage at which the MIT occurs. The resistivity increases by only about two orders of magnitude reaching $\rho_{2.15\ dpa,\ 300K}$ = 52 mΩ cm, and the MIT appears at a lower dose of 0.05 dpa. Notably, the electrical resistivity saturation at 52 mΩ cm is reached at 1.4 dpa. Similarly as the films on MgO, the film progressively undergo a transition from metallic to semiconductor, from 0.05 dpa (FIG 4(c), point B) to 0.01 dpa (FIG 4(d), point C). However, in comparison with the film deposited on MgO, the film deposited on $Al_2O_3$ seems to be more sensitive to implanted defect as the film becomes completely semiconductor in the measurement temperature range at only 0.2 dpa (FIG 4(d), point C). Therefore, the as-deposited state of the films plays a predominant role in the evolution of both resistivity and conduction modes during defect accumulation induced by ion irradiation.

Interestingly, when the material is in second regime (regime II, cumulatively implanted at 2.15 dpa), a post-implantation annealing 873 K restores the electrical transport characteristic similar to the level of a film in the first regime (Regime I) (see FIG 4(a), the dotted arrows). This observation indicates that the two regimes (regime I below 0.8 dpa in the MgO case and 0.1 dpa for $Al_2O_3$ and the regime II above) originate from distinct defect-creation processes: the defects formed in regime II are comparatively unstable and can be annealed out at relatively low temperature (≈750 K), whereas those generated in regime I are more stable and persist after temperature annealing, at least up to 873 K. For comparison, another film on $Al_2O_3$ was implanted directly at 0.4 dpa and then annealed at 750 K (represented by the stars FIG 4(a)). The same behavior is observed as the cumulatively implanted film.

### C. Damage accumulation models

Several phenomenological models have been developed to describe the accumulation of defects in irradiated materials, with the extent of damage generally assessed using structural characterization techniques such as X-ray diffraction (XRD) and Rutherford backscattering spectrometry (RBS). The Gibbons' model describes the fraction of damage $f_a$, resulting either from direct defect formation (called DI: direct impact (n=1)) along a single ion track or from the overlap of two or more tracks (n=2 and more): [38]

*charlotte.poterie@univ-poitiers.fr
**arnaud.lefebvrier@kemi.uu.se

TABLE II. Values for the fits of equation 6 (with m=2) to experimental resistivity data at 300 K displayed in FIG 3(a) and FIG 4(a). n corresponds to the number of ion impacts required to create permanent damage to the material for each step and their respective saturation resistivity $\rho_i^{sat}$ and cross-section $\sigma_i$. $\Phi_c$ denotes the dose interval at which the MIT occurs.

| | **Regime I** | | | $\Phi_c$ (dpa) | **Regime II** | | | **Fit results** | |
|---|---|---|---|---|---|---|---|---|---|
| | n | $\rho_1^{sat}$ (mΩ cm) | $\sigma_1$ (dpa$^{-1}$) | | n | $\rho_2^{sat}$ (mΩ cm) | $\sigma_2$ (dpa$^{-1}$) | $R^2$ | RMSE (mΩ cm) |
| **MgO** | 1 | 8.5 ± 0.5 | 0.34 ± 0.02 | 0.8 - 1.15 | 7-10+ | 700 ± 150 | 3.5 ± 1 | 0.975 | 11 |
| **Al$_2$O$_3$** | | 1.5 ± 0.5 | 13 ± 5 | 0.05 - 0.15 | | 52 ± 2 | 9 ± 1.5 | 0.99 | 1.8 |

$$f_a = f_a^{sat}\left[1 - \sum_{k}^{n-1}\frac{(\sigma\phi)^k}{k!}\exp(-\sigma\phi)\right] \quad (4)$$

where $f_a^{sat}$ is the value at saturation, $\sigma$ the disordering effective cross section, $\phi$ the ion fluences (or dpa) and $n$ the number of ion impacts required to create permanent damage to the material. When defect accumulation occurs in multiple steps or regimes, the Multi-Step Damage Accumulation (MSDA) model has been proposed: [39]

$$f_a = \sum_{i=1}^{m}(f_{a,i}^{sat} - f_{a,i-1}^{sat})G\{1 - exp[-\sigma_i(\phi - \phi_i)]\} \quad (5)$$

where $m$ is the number of steps required for the achievement of the total process, $G$ is a function which transforms negative values into 0 and leaves positive values unchanged. Unlike the model by Gibbons, the MSDA model considers discontinuous steps where each step is regarded as only Direct Impact (DI) process ($n$=1).

In the present work, these models provide the framework for our analysis, under the assumption that the dose-dependent evolution of the electrical resistivity is a direct and reliable probe of defect accumulation during irradiation. On this basis, we propose the following expression, which accounts for the distinct resistivity regimes observed for films grown on both substrates (at T = 300 K; see FIG 2(a) and FIG 3(a)):

$$\rho(\phi) = \rho_{ref} + \sum_{i=1}^{m}\left\{\rho_i^{sat}\left[1 - \sum_{k}^{n-1}\frac{(\sigma_i\phi)^k}{k!}\exp(-\sigma_i\phi)\right]\right\} \quad (6)$$

FIG 2(a) and 3(a) demonstrate that Eq. (6) quantitatively captures the evolution of resistivity with fluence in ScN using two distinct steps (m=2). Table 2 summarizes the extracted model parameters. Because the saturation resistivity values, $\rho_i^{sat}$, were difficult to determine precisely, the fits were optimized based on a combined assessment of the coefficient of determination ($R^2$) and the root-mean-square error (RMSE), which are also reported in Table 2.

Regime I, occurring at low damage doses, corresponds to a Direct Impact (DI) process (Eq. (6), with $n = 1$), in which a single ion impact is sufficient to create a permanent defect or an electrically active scattering center. This resulting population of defects $N_D(dpa)$ gives rise to the additional resistivity term $\rho_D(dpa)$ introduced in Eq. (1) and also participate to carrier scattering reducing $\mu_D(dpa)$ in Eq. (2), $1/\mu_D(dpa) \propto N_D(dpa)$. [40] Regime II appears at higher damage doses and reflects a process of damage accumulation (DA), i.e. the progressive accumulation of damage in a same volume, during which additional ion impacts increasingly interact with pre-existing damaged regions. The extracted value of $n$ (Eq. (6), $n \geq 7$) indicates that the resistivity is resilient to single impact and increases significantly when multi-hit accumulates. This regime likely involves the formation of new electrically active defects, consistent with the partial resistivity recovery observed upon annealing (see FIG 4(a)). At the same time, the semiconducting behavior becomes predominant, confirming that charge carriers are becoming more localized due to increased disorder. In this semiconducting regime, the defect density is high, leading to a carrier mobility that remains roughly constant with temperature, resulting in the $\mu_D$ term which dominates in Eq. (2). For example, at the maximum of 2.15 dpa, the mobility is $\mu(T) \approx \mu_{D, 2.15dpa, 300 K} \approx 1.2(2)$ cm² V⁻¹ s⁻¹ for the film on the Al$_2$O$_3$ substrate. The efficiency of electrically active defect creation depends strongly on the substrate, i.e. on the initial quality of the film.

*charlotte.poterie@univ-poitiers.fr
**arnaud.lefebvrier@kemi.uu.se

In the first regime, the ScN/MgO sample exhibits greater irradiation resistance, with respect to electrical properties, as evidenced by a significantly smaller damage cross section ($\sigma_1$) compared with the film grown on $Al_2O_3$ (almost 40 times smaller). Consequently, for the $Al_2O_3$-supported film, the MIT transition occurs at lower doses. In Regime II, the damage cross sections ($\sigma_2$) are of the same magnitude which highlights that the cascades inducing damage accumulation interact mainly with the pre-existing damaged region rather than with the residual state. Both films have reached their disordering saturation which results in the MIT. Hence, the multi-hit process occurs on analogous disordered region which might explain the similar cross sections ($\sigma_2$). However, the residual state has an effect on the accommodation of the defect density. Unlike in the ScN/MgO case, the saturation of the electrical resistivity becomes apparent above 1.4 dpa in the ScN/$Al_2O_3$ case. Upon further damage accumulation (above ~ 2 dpa), the resistivity even slightly begins to decrease ($\rho_{1.7dpa}$ = 52 mΩ cm to $\rho_{2.15dpa}$ = 50 mΩ cm), deviating from the expected monotonic increase typically associated with defect buildup. Such a decrease suggests that the film relaxes to lower its system energy, as observed in some cumulatively irradiated materials. [41,42] In the case of ScN/MgO, no saturation nor decrease of the electrical resistivity is observed which might indicates that the system withstand higher deformation. For both substrates, this second step coincides with the onset and development of the metal–insulator transition (MIT), confirming that the accumulated disorder not only increases scattering but also progressively localizes charge carriers. The differences observed between the two substrates – specifically, the earlier onset of the second step and the lower saturation resistivity for ScN on $Al_2O_3$ are directly related to the different initial defect landscapes and structural states of the reference films.

## IV. DISCUSSION:
## Carrier localization and MIT

The resistivity-accumulation model provides clear insight into the disordering rate and underline the influence of the substrate on the initial film quality. An additional and essential consequence of irradiation is carrier localization. Localization in ScN thin films has been already reported under ion implantation, [1,3,19–21] despite variations in substrates and resulting microstructures. To confirm that the second regime observed in our measurements is indeed associated with the generation of localized states, a ScN/$Al_2O_3$ film was irradiated to 0.4 dpa and subsequently annealed at 750 K (represented by a star in FIG 4(a)). As shown in FIG 5., carrier localization disappears after low-temperature annealing, leaving only defects generated by direct ballistic impacts (n = 1, regime I). Similarly, annealing at 873 K (600°C) of the film implanted at 2.15 dpa (see FIG 4(a)) further confirms the key role of disorder in driving localization. After annealing, the resistivity returns to values characteristic of regime I, indicating that the defects responsible for regime II are largely unstable and recover upon thermal treatment. In this work, ion implantation is carried out at high energies, and the implanted oxygen ions pass entirely through the film; thus, only intrinsic defects are produced such as vacancies ($V_N$ and $V_{Sc}$), interstitials ($X_i$) or complexes ($V_y$-X). FIG 6. shows the RT variations of Hall carrier concentration versus damage accumulation (dpa). The carrier concentration shows an overall decrease over the dpa range except for a small increase observed at low damage regime in the ScN/Al$_2$O$_3$ case (see inset FIG 6.). It is clear that deep acceptor traps participate in the defect physics relevant to electrical degradation. The carrier concentration is divided by about 10 at doses larger than 1.5 dpa. Interestingly is that again the quality of the film is underlined at low doses where the film on Al$_2$O$_3$ substrate show a first stage of carrier generation suggesting an activation of preexisting donor defects or a partial relaxation of strain. Overall, there remains a lack of a comprehensive defect database for ScN. Nevertheless, existing studies consistently identify the Sc vacancy $V_{Sc}$ as an acceptor-type defect. [10,43] During irradiation, additional defect complexes such as ($V_N$-$X_i$) or ($V_{Sc}$-$X_N$), with $X_i$ interstitials impurities and $X_N$ substitution of N by impurities, can also form, introducing deep acceptor levels in the bandgap and acting as efficient scattering centers for carriers. Among these, the ($V_{Sc}$-$O_N$) complexes with large binding energy have been theoretically shown to behave as strong acceptors [14,20] thereby reducing the free-electron concentration and degrading mobility. [14] This is consistent with experimental observations reporting mobility loss in O- and C-implanted ScN. [19,20] It is therefore important to note that our as-deposited layers already contain ~1% oxygen, which is responsible for their n-type degenerate behavior and provides a reservoir for the formation of such vacancy–oxygen complexes during irradiation.

*charlotte.poterie@univ-poitiers.fr
**arnaud.lefebvrier@kemi.uu.se

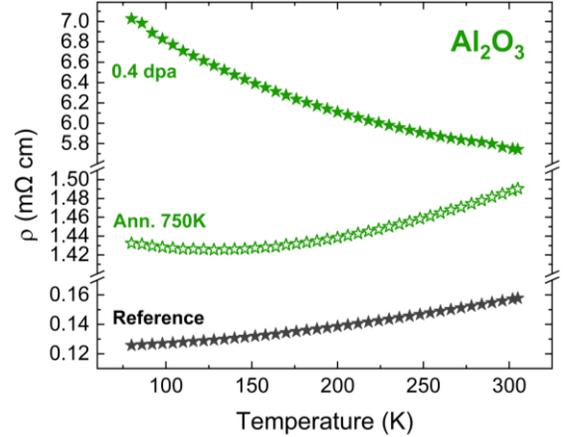

FIG 5. Temperature-dependent resistivity of the ScN film directly implanted at 0.4 dpa, exhibiting a negative temperature coefficient (dρ/dT < 0), and after in situ annealing at 750 K. The recovery of a positive temperature coefficient after annealing indicates the complete suppression of carrier localization and the restoration of regime I transport behavior.

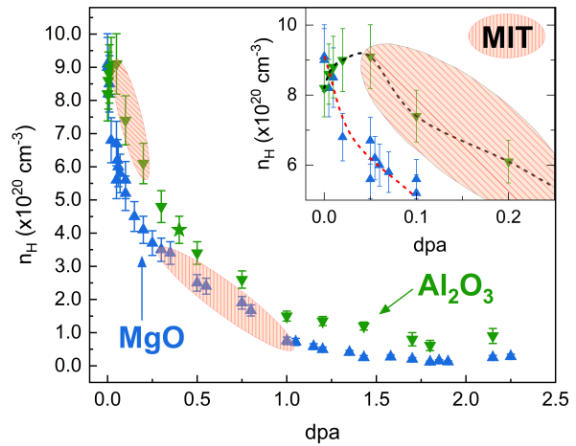

FIG 4. Evolution of the Hall carrier concentration $n_H$ with accumulated damage (dpa). The red hatched areas highlight the dose dependent MIT interval. The inset emphasizes the early-stage increase in $n_H$ observed at low doses in the ScN/$Al_2O_3$ film.

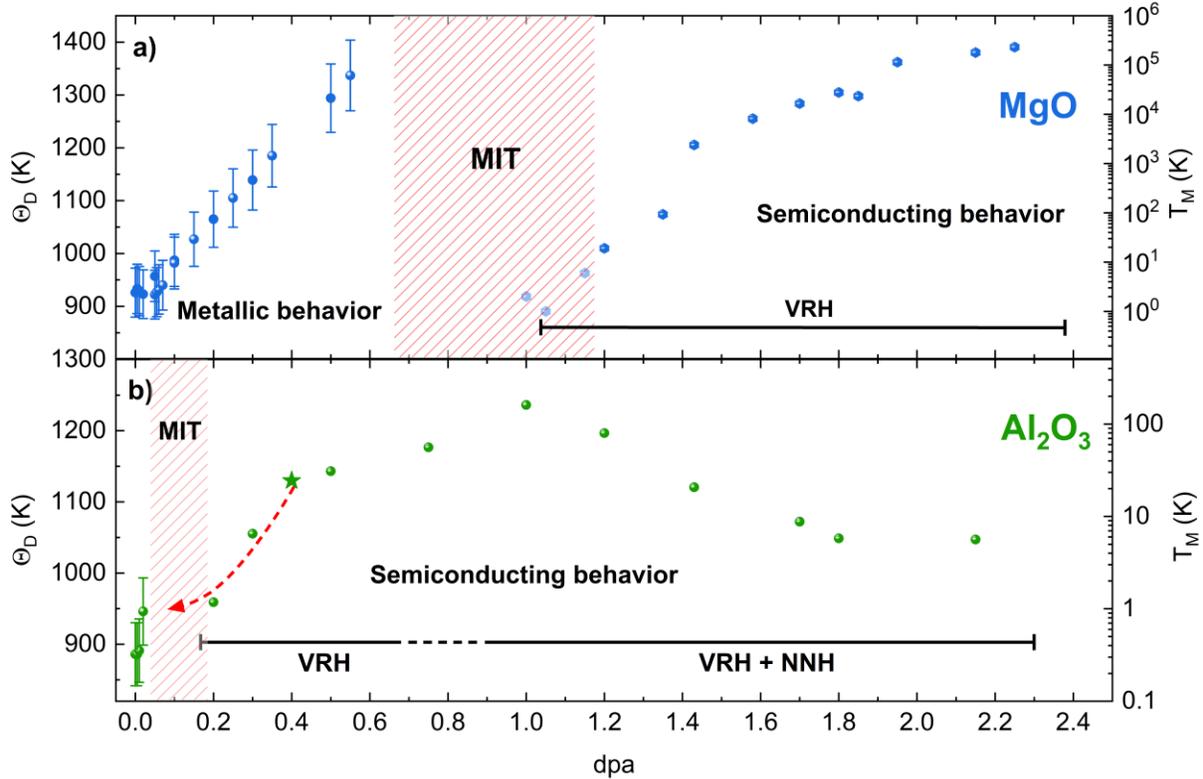

FIG 6. Overview of the evolution the Debye temperature $\Theta_D$ (Eq. 2) and of the Mott temperature $T_M$ of ScN under O-irradiation for both substrates. The red hatched zone represents the MIT over the damage interval. The $T_M$ temperatures are determined according to Eq.7 by plotting $\ln\rho(T)$ vs. $T^{1/4}$. Above 1 dpa for ScN/Al$_2$O$_3$, a combination of VRH and NNH is observed. The red dashed arrow highlights the restored metallic conduction induced by defect recovery resulting from *in situ* annealing at 750 K of the film implanted at 0.4 dpa, as seen FIG 2.

FIG 7. provides an overview of the evolution of the characteristic temperatures ($T_M$ and $\Theta_D$) describing the

*charlotte.poterie@univ-poitiers.fr
**arnaud.lefebvrier@kemi.uu.se

electrical properties of ScN under irradiation for both substrates. At low doses, below the MIT, the films remain metallic and the primary effect of irradiation is the increase in the additional resistivity $\rho_D(dpa)$, which reflects the progressive buildup of defects inducing increased scattering. As damage accumulates, the crystalline lattice becomes increasingly perturbed, leading to a rise in compressive stress within the films. [2,30] This stress stiffening results in an increase of the Debye temperature $\theta_D$, consistent with a reduction in phonon vibrational amplitudes and a progressive deviation from the pristine metallic state. Considering the film on MgO, at 0.1 dpa, thus in the low-dose regime, where irradiation introduces temperature-independent contributions to the resistivity and mobility ($\rho_D$, $\mu_D$) without altering their temperature dependence, it is possible to calculate the diffusion centers density $N_D$. At 300 K, the carrier concentration decreases from $n_{ref}=9.0\times10^{20}$ to $n_{irr}=5.8\times10^{20}$ $cm^{-3}$, yielding a trapped carrier density $n_{trap}=3.1\times10^{20}$ $cm^{-3}$ ($\approx 35\%$). SRIM simulations predict a vacancy density of $8\times10^{21}$ $cm^{-3}$ at 0.1 dpa, implying that only ~4% of the created vacancies remain electrically active. The density of irradiation-induced diffusion centers was estimated from the mobility degradation. In a degenerate semiconductor, the defect-related momentum-relaxation rate is given by $(1/\tau_D)=N_D\sigma_D v_F$, [40] with $(1/\tau_D)=(e/m*)(1/\mu_{irr}-1/\mu_{ref})$. Using $m*=0.5m_0$ [26], $v_F=5\times10^5$ $m\ s^{-1}$ (for $E_F-E_C=0.2\ eV$ [26]), and $\sigma_D=10^{-19}\ m^2$, the diffusion-center density at 0.1 dpa is estimated as $N_D\approx2\times10^{21}\ cm^{-3}$. This hierarchy $N_D>>n_{Hall}>n_{trap}$ demonstrates that irradiation predominantly generates weakly active defect complexes that are highly efficient momentum-relaxing centers. This decoupling between carrier trapping and elastic scattering provides direct evidence that, in irradiated ScN, charge transport at low dose is governed by disorder-driven metallic conduction rather than by carrier depletion.

Once the MIT is reached, the transport properties become governed by defect-induced carrier localization. In this regime, hopping transport progressively dominates, in particular the Mott Variable Range Hopping (Mott-VRH). [3,19–21] This disorder induced localization can be characterized by the Mott temperature $T_M$. [44] This temperature is equivalent to a characteristic energy barrier quantifying the capacity of electrons to hop between localized states. $T_M$ can be extracted through the fit of the resistivity by a Mott-VRH mechanism (p=1/4 in 3D conduction) given by:

$$\rho(T) = \rho_0 exp\left[\left(\frac{E_t}{k_B T}\right)^p\right] = \rho_0 exp\left[\left(\frac{T_M}{T}\right)^{\frac{1}{4}}\right] \quad (7)$$

In the case of Mott-VRH, Coulomb interactions between electrons are neglected which leads to $T_M$ function of the localization length $\xi$ and a constant density of state near the Fermi level $N(E_F)$: $T_M = \frac{18}{\xi^3 N(E_F) k_B}$. With increasing damage accumulation, the Mott temperature is increasing which signifies that the localization length is decreasing reflecting the defect build-up with increasing dose. As seen on FIG 7., the film on MgO shows high values of $T_M$, $T_M > 10^5$ K at 2.25 dpa highlighting that the film is highly resistive emphasizing a very strong localization. In comparison, the film on $Al_2O_3$ exhibits a maximum $T_M$ at 1 dpa, of ~250 K and is then decreasing as defects are further accumulating. The much lower $T_M$ compared with MgO indicates that electrons are less localized. With increasing dose, we observe some deviation on the VRH mechanism especially toward the high temperature of resistivity curves suggesting an apparition of another conduction mode, likely through Nearest Neighbor Hopping (NHH) conduction. Hence, with increase hit overlap, the conduction on ScN/$Al_2O_3$ in regime II can be expressed as:

$$\frac{1}{\rho_{reg\ II}(T)} = \sigma_{VRH}(T) + \sigma_{NNH}(T) \quad (8)$$

This could be consistent with resistivity measurements showing saturation at high doses suggesting another mechanism of decreasing $T_M$, as the defect overlapping to create different conduction path or the formation of more extended defects such as dislocations clusters relaxing the disorder. [23]

Finally, using a damage-accumulation model on the resistivity, we establish a clear hierarchy in defect formation mechanisms in ScN films with different initial defect landscapes. In the early stage of irradiation, when a fraction of Frenkel pairs survives dynamical recombination, stable acceptor-type complex defects of the form (V–X) are preferentially formed, where X corresponds to unintentional impurities incorporated during growth (e.g., oxygen). The formation of these complexes rapidly saturates due to the limited mobility of the involved species. Further irradiation does not simply increase their concentration but instead drives the accumulation of additional point defects that can only stabilize within an increasingly disordered local environment. This second stage requires significant hit overlap (regime II), corresponding to a heavily damaged volume and marking the onset of defect-induced carrier localization. Importantly, the point defects responsible for localization are comparatively weakly bound and are efficiently annihilated by moderate thermal annealing (≈500–600 °C), leading to a full recovery of metallic transport. This behavior unambiguously demonstrates that localization in irradiated ScN is governed by disorder-driven, metastable defect configurations rather than by permanent chemical doping.

*charlotte.poterie@univ-poitiers.fr
**arnaud.lefebvrier@kemi.uu.se

## V. CONCLUSIONS

This study explored how electrical transport in ScN films evolves with accumulated irradiation damage and showed that the substrate (MgO or $Al_2O_3$) strongly influences both the initial defect landscape and the way the films respond to increasing disorder. The resistivity evolution revealed two distinct regimes associated with a metal-to-semiconductor transition, which occurs when defect-induced localized states dominate and hopping conduction replaces metallic transport. We proposed an adapted phenomenological model that quantitatively reveals that two distinct defect-generation processes are involved. At low doses (regime I), the damage accumulation is well described by the direct-impact (DI) model, which manifests as an additional resistivity term associated with stable isolated complex defects of the form (V–X), involving residual impurities X such as oxygen. At higher doses (regime II), a multi-hit process leads to a second disorder regime dominated by carrier localization induced by the stabilization of point defects such as Sc vacancies. These defects are all predominantly acceptor-type defects, reducing the free-electron concentration and mobility. The critical damage level and localization strength are strongly substrate dependent, underscoring the key role of the initial residual defect state.

## ACKNOWLEDGMENTS

The authors acknowledge funding from the Swedish Government Strategic Research Area in Materials Science on Functional Materials at Linköping University (Faculty Grant SFO-Mat-LiU No. 2009 00971), the Knut and Alice Wallenberg foundation through the Wallenberg Academy Fellows program (KAW-2020.0196), the Swedish Research Council (VR) under Project Nos. 2021-03826 (P. E.), 2025-03680 (P. E.), and 2025-03760 (A. F.), and the Swedish Energy Agency under project number 52740-1. This work was also supported by the French government program "Investissements d'Avenir" (EUR INTREE – reference ANR-18-EURE-0010, LABEX INTERACTIFS – reference ANR-11-LABEX-0017-01, and UP-SQUARED – reference ANR-21-EXES-0013).

*charlotte.poterie@univ-poitiers.fr
**arnaud.lefebvrier@kemi.uu.se

*charlotte.poterie@univ-poitiers.fr
**arnaud.lefebvrier@kemi.uu.se

*charlotte.poterie@univ-poitiers.fr
**arnaud.lefebvrier@kemi.uu.se